\documentclass[pre,aps,12pt,amssymb,bibnotes,footinbib]{revtex4}
\usepackage{graphicx}
\usepackage{dcolumn}
\usepackage{bm}

\begin{document}

\title{Temperature dependence of circular DNA topological states}% Force line breaks with \\

\author{Hu Chen$^{1,\dagger}$}
\author{Yanhui Liu$^{2,3,1,\dagger}$}
\author{Zhen Zhou$^1$}
\author{Lin Hu$^2$}
\author{Zhong-Can Ou-Yang$^3$}
\author{Jie Yan$^1$}\email[Email address: ]{phyyj@nus.edu.sg}
\affiliation{$^1$ National University of Singapore, Department of Physics, 2 Science Drive 3, Singapore, 117542 \\
$^2$ Guizhou Provincial Key Lab for Photoelectron Technology and Application, Guizhou University, Guiyang, 550025 \\
$^3$ The Institute of Theoretical Physics, Chinese Academy of Sciences, Beijing, China, 100080
}
\footnotetext[1]{$\dagger$ These authors contributed equally to this work.}
\date{\today}% It is always \today, today,

\begin{abstract}
Circular double stranded DNA has different topological states which are defined by their linking numbers. Equilibrium distribution of linking numbers can be obtained by closing a linear DNA into a circle by ligase. Using Monte Carlo simulation, we predict the temperature dependence of the linking number distribution of small circular DNAs. Our predictions are based on flexible defect excitations resulted from local melting or unstacking of DNA base pairs. We found that the reduced bending rigidity alone can lead to measurable changes of the variance of linking number distribution of short circular DNAs. If the defect is accompanied by local unwinding, the effect becomes much more prominent. The predictions can be easily investigated in experiments, providing a new method to study the micromechanics of sharply bent DNAs and the thermal stability of specific DNA sequences. Furthermore, the predictions are directly applicable to the studies of binding of DNA distorting proteins that can locally reduce DNA rigidity, form DNA kinks, or introduce local unwinding.
\end{abstract}

% PACS, the Physics and Astronomy
\maketitle

\newpage

\section{Introduction}
%\definecolor{Red}{rgb}{1,0,0}

In a closed circular double stranded DNA molecule, the two strands are linked. The number of times the two strands wind around each other is a topological invariant, and it is called the linking number $\textrm{Lk}$, an integer that may be positive or negative depending on the orientation of the two strands. Due to the right-handed double helical structure of DNA, there is an intrinsic linking number $\textrm{Lk}_0 = N/\gamma$, where $N$ is the number of base pairs (bp) of the DNA, and $\gamma = 10.5$ is the number of base pairs per helical turn of DNA. It is more convenient to describe the topological states of the circular DNA using the linking number difference $\Delta \textrm{Lk} = \textrm{Lk} - \textrm{Lk}_0$. Since $\textrm{Lk}_0$ is not an integer, $\Delta \textrm{Lk}$ is not an integer either. However, $\Delta \textrm{Lk}$ can only defer by integer numbers. Previous experiments and simulations show that, the equilibrium distributions of $\Delta \textrm{Lk}$ for circular DNA in the range of 250 - 10,000 bp may be approximated by Gaussian distributions. A thorough review of linking number distributions of circular DNA can be found in \cite{Vologodskaia2002}. 

It is well known that the linking number puts a constraint on the twist and the writhe of a DNA \cite{White1969,Fuller1971}:
\begin{equation}
 \Delta \textrm{Tw} + \textrm{Wr} = \Delta \textrm{Lk},
\end{equation}
where $\Delta \textrm{Tw}$ is the difference between the actual twist of the circular DNA and the relaxed twist of the same DNA of linear form; $\textrm{Wr}$ is the writhe of the DNA backbone. Due to this constraint, the distribution of $\Delta \textrm{Lk}$ may be obtained from convolution of the distributions of $\Delta \textrm{Tw}$ and $\textrm{Wr}$ \cite{Podtelezhnikov1999}:
\begin{equation}\label{eq.lk}
 \rho(\Delta \textrm{Lk}) = \int d \textrm{Wr} \rho_{tw}(\Delta \textrm{Lk} - \textrm{Wr}) \rho_{wr}(\textrm{Wr}).
\end{equation}
Please note that an assumption is made where $\rho_{tw}(\Delta \textrm{Tw})$ and $\rho_{wr}(\Delta \textrm{Wr})$ are independent, which is a reasonable approximation according to \cite{Vologodskii1993}. 

For a linear DNA of a contour length $L$, its twist energy in $k_B T$ unit can be approximated by an harmonic form: $E(\Delta \textrm{Tw}) = 2\pi^2 (C/L) \Delta \textrm{Tw}^2$, where $C=75$ nm is the canonical twist persistence length of a B-DNA \cite{Vologodskaia2002,Moroz1998,Bryant2003,Shore1983}. The harmonic form of the twist energy leads to a Gaussian distribution of the twist with a variance of $\sigma_{tw}^2 = {L\over 4\pi^2 C}$. On the other hand, the writhe distribution depends on the bending elasticity model of DNA. In a discretized DNA model, a DNA can be considered to be a chain consisting of $N$ straight segments of a segment length $b\ll A$, where $A=50$ nm is the persistence length of linear B-form DNA. 
%In this paper, we choose $b= 1\text{ nm}=3\text{ bp}$. 
The bending energy is carried by the $N$ vertices in a circular DNA:
\begin{equation}\label{eq.e}
E=\sum_{i=1}^{N}E_i(\hat{t}_i, \hat{t}_{i+1}),
\end{equation}
where $E_i$ is the bending energy of the $i^{th}$ vertex connecting the two adjacent tangent vectors $\hat{t}_i$ and $\hat{t}_{i+1}$. 

In the worm-like chain (WLC) model of DNA, $ E_i = \frac{A}{2b}(\hat{t}_{i+1}-\hat{t}_i)^2$, where the persistence length $A$ is also the correlation length of local tangent vectors \cite{Smith1992,Marko1995,Bustamante1994,Shore1983}. Previously, Eq. (\ref{eq.lk}) was used to obtain the linking number distributions based on the WLC model \cite{Podtelezhnikov1999}. The predicted $\rho(\Delta \textrm{Lk})$ were found in good agreement with those measured in experiments \cite{Vologodskaia2002} for circular DNAs of 250 - 10,000 bp sized in a wide temperature range of 4 - 37 degree Celsius \cite{Horowitz1984,Depew1975,Rybenkov1997}. In addition, the WLC model has been supported by numerous other experiments including the force-extension measurements of single DNAs of micrometer(s) long \cite{Smith1992,Marko1995,Bustamante1994}, and the ``j-factor'' measurement to measure the looping probability of DNA of larger than 230 bp \cite{Shore1983}. It has no doubt that for DNA larger than $200$ bp, at temperature below 37 degree Celsius, the WLC model with choice of $A$ in 45-53 nm is suitable to describe the DNA bending elasticity. 

However, the above experiments and simulations do not rule out the possibility that WLC may fail to describe DNA elasticity at sharper bending conditions (for example, small DNA loops), or/and at higher temperatures. The most likely defects are melted base pairs or base pairs that lose their stacking interactions with its neighboring base pairs. The possibility of excitation of such defect was first pointed out by Crick and Klug \cite{Crick1975}, and recently was observed in molecular dynamics simulation \cite{Lankas2006}, where the defect is an unstacked base-pair step that allows formation of $> 90$ degrees of bending at the defect site. Furthermore, such defects were also observed in $\leq 65$ bp minicircles recently \cite{Du2008}. We note that both melting and unstacking are sensitive to temperature. Theoretically, it has been shown that rare excitations of flexible defects may have drastic effects on the looping efficiency of small DNA \cite{Yan2004,Wiggins2005,Du2005}. Such sensitivity to excitations of flexible defects is expected to exist in various other aspects of micromechanics of sharply bent DNAs. In this research, the effects of the flexible defect excitation on $\rho(\Delta \textrm{Lk})$ of small circular DNAs are investigated based on an assumption that the defects have a reduced bending rigidity. The additional effects of possible reduced twist rigidity and unwinding of defect are also considered.

\section{Effects of the defect flexibility on linking number distribution}

A defect may change various local mechanical properties, such as bending rigidity, twist rigidity, and may lead to unwinding of DNA. In this section, we show that the reduced bending rigidity alone is sufficient to cause measurable increase of variance of the linking number distribution for small circular DNAs. 
A DNA bending energy model that includes the excitations of flexible defects should contain three critical parameters: the persistence length of B-DNA $A$, the persistence length of the defect $A'$ ($A'\ll A$), and the energy required to excite one such defect $\mu$ (in $k_B T$ unit). As shown in \cite{Yan2004}, the vertex energy can be straightforwardly written as:
\begin{equation} \label{WLC_bubble}
E_i = {\left(\delta_{n_i,0} a + \delta_{n_i,1} a'\right) \over 2} \left(
\hat{\bf t}_{i+1}-\hat{\bf t}_i \right)^2 + \mu \delta_{n_i,1},
\end{equation}
where the $n_i$ are two-state variables, indicating whether segment $i$ is either in double helix form ($n_i=0$) or contains a defect ($n_i=1$); $a=A/b$ and $a'=A'/b$ are the vertex bending rigidity of the regular DNA site and the defected DNA site, respectively. In our simulation, $A=50$ nm is the B-DNA persistence length, $A'=1$ nm is close to the bending persistence of single stranded DNA, and $\mu$ is scanned in the range 4 - $\infty$ $k_B T$. The segment length $b$ is chosen to be $1$ nm $=3$ bp. This does not mean the defect size has to be 3 bp large. Instead, it means the defect may have a size from 1 bp up to 3 bp, so it includes the defect at single base pair step level as suggested in \cite{Crick1975,Lankas2006}. As such, $A'$ in our model is an average rigidity over all possible defects, and $\mu$ is the average excitation energy.

In the above model, the Ising degree of freedom $\{n_1, n_2,\cdots, n_N\}$ is independent of the conformational degree of freedom $\{\hat{t}_1, \hat{t}_2,\cdots,\hat{t}_N\}$, so an effective vertex energy that only depends on the conformation may be derived by summation over the Ising degree of freedom, as shown in \cite{Yan2005}:
\begin{equation}\label{energy}
E_i = -ln[e^{-{a\over 2}(\hat{t}_{i+1}-\hat{t}_i)^2} +
e^{-\mu-{a'\over 2}(\hat{t}_{i+1}-\hat{t}_i)^2} ].
\end{equation}
Since Eq. (\ref{energy}) only depends on conformation, it is more convenient for Monte Carlo (MC) simulation. Hence, it was used in our simulation for writhe distribution of circular DNAs. A vertex is determined to be defected when its bending angle exceeds a threshold bending angle $\theta_c = \arccos (1-{\mu \over a-a'})$ (please refer to Eq. (23) of section E.3 in \cite{Yan2005}).

Based on Eq. (\ref{energy}), MC simulation is used to sample conformations of circular DNA. To update the conformation, a subchain is rotated by a random angle around the straight line connecting two randomly chosen vertices. Metropolis criterion is applied to determine whether the new conformation is accepted. During the simulation, the distribution of writhe number is obtained. For each simulated DNA circle, its writhe is computed by
 \cite{Hao1989,Sucato2004}:
\begin{equation}
\textrm{Wr} = {b^2\over 4 \pi} \sum\limits_{i,j=1;i\neq j}^{N} {\hat{t_j}
\times \hat{e}_{ij}\cdot \hat{t}_i \over d_{ij}^2},
\end{equation}
where $\hat{e}_{ij}$ is the unit vector pointing from the $j^{th}$ vertex to the $i^{th}$ vertex, and $d_{ij}$ is the distance between the two vertices. 
Normalized histogram of the simulated writhe numbers being divided by the bin size gives the density distribution function $\rho_{wr}(\textrm{Wr})$. The volume exclusion is included in the simulation in which conformations with $d_{ij}<3.4$ nm are not accepted. 

\begin{figure}
\includegraphics[width=6cm]{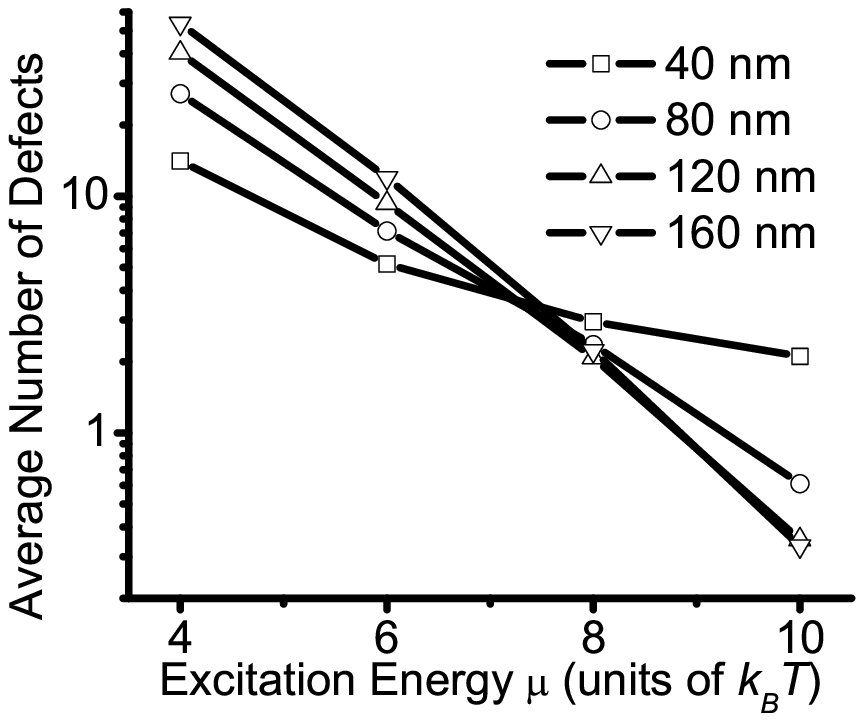}
\caption{The average number of defects in circular DNA as a function of the excitation energy $\mu$.
}
\label{fig_num}
\end{figure}

%chen: number of defects in the simulation
Fig. \ref{fig_num} shows the average number of defects for circular DNA from 40 nm to 160 nm. For $\mu=10$ $k_B T$, there are $<1$ defects in circular DNA bigger than 80 nm, while there are about 2 defects in small circular DNA of 40 nm. This is consistent with the result obtained previously by transfer matrix calculation \cite{Yan2005}. When $\mu$ decreases to 6 or even 4 $k_B T$, the average number of defects begins to increase with increasing DNA length. At 4 $k_B T$, about 1/3 of the DNA is melted.

\begin{figure}
\includegraphics[width=6cm]{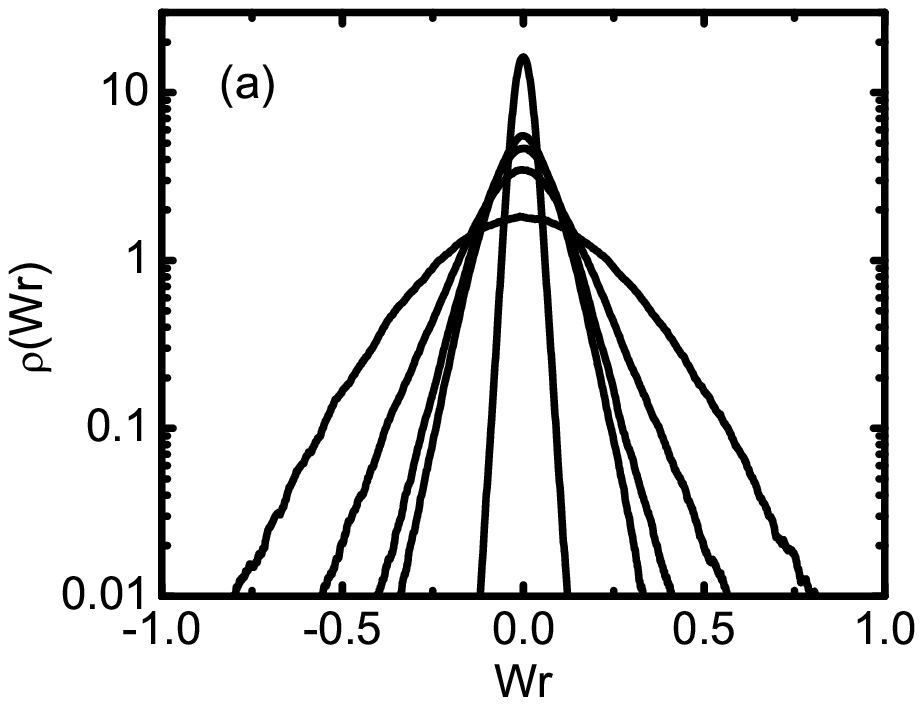}% Here is how to import EPS art
\includegraphics[width=6cm]{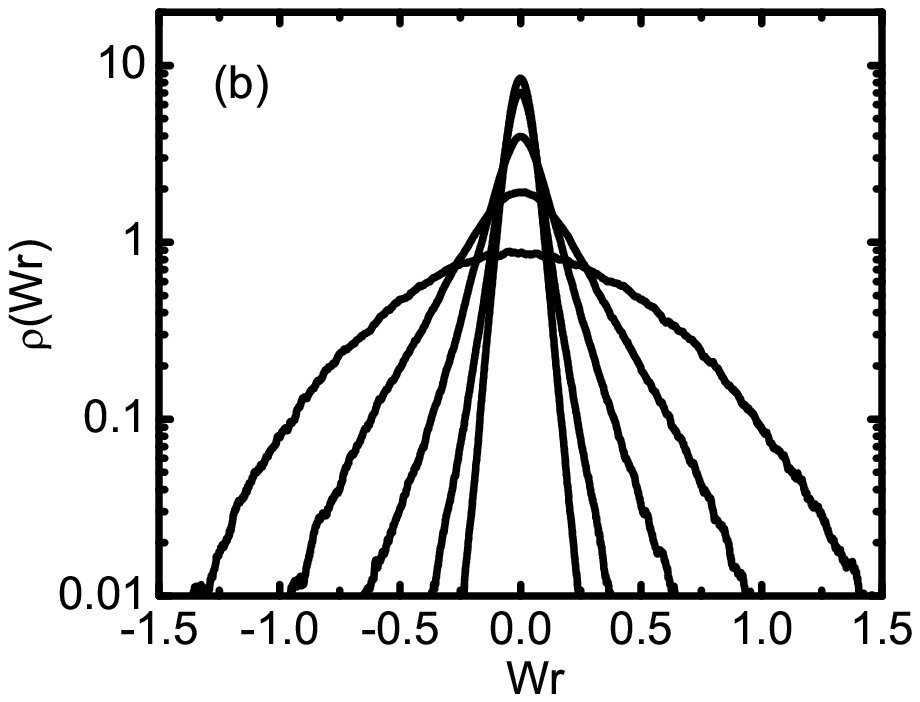}
\includegraphics[width=6cm]{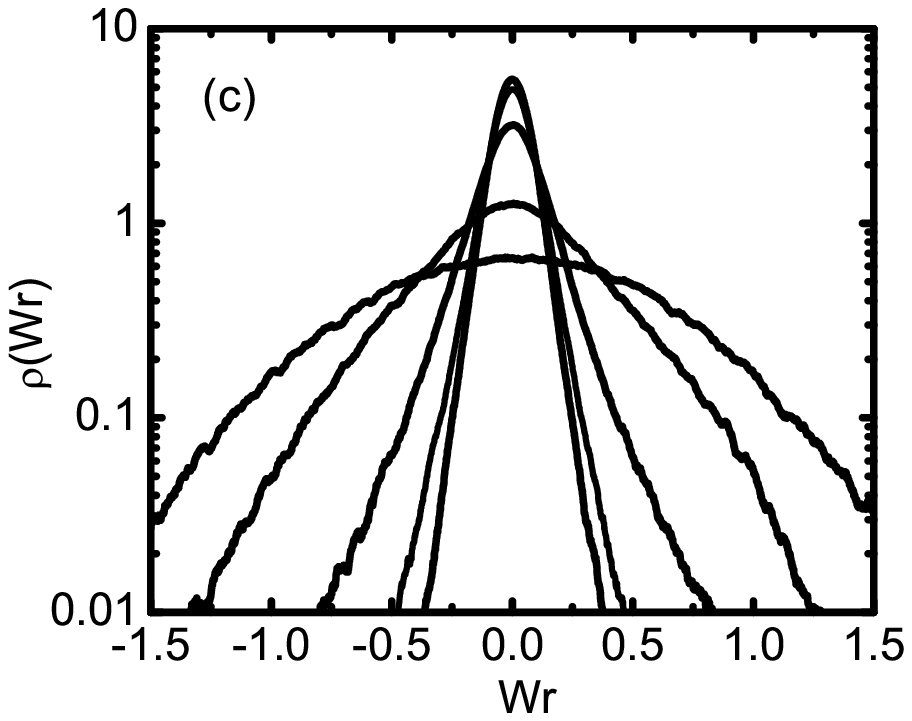}
\includegraphics[width=6cm]{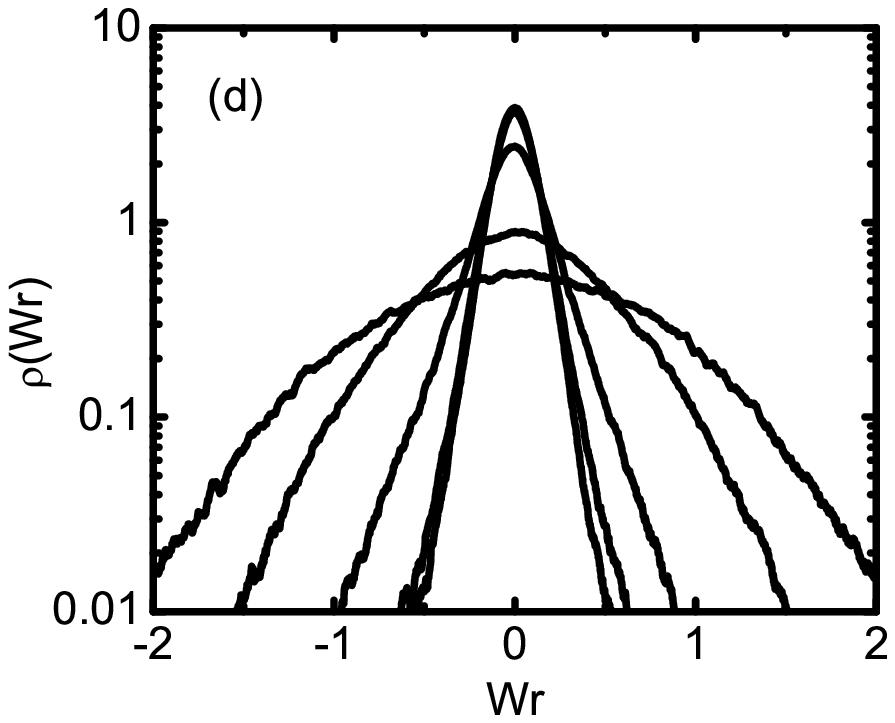}
\caption{Writhe number distributions based on Eq. (\ref{energy}) of DNA minicircles of sizes $40$ nm (a), $80$ nm (b), $120$ nm (c), and $160$ nm (d), respectively. For each figure, from outside to inside, the excitation energies are $4$, $6$, $8$, $10$ $k_B T$, and $\infty$ (B-DNA), respectively.
}
\label{fig_writhe}
\end{figure}

In this section, we only consider the effects of the reduced bending rigidity of the defects, so we set the twist rigidity of the defect to be unchanged: $C' = C = 75$ nm. Therefore, the defect excitation does not affect the twist distribution: $\rho_{tw}(\Delta \textrm{Tw})$ is a Gaussian with a variance: $\sigma_{tw}^2 ={L\over 4 \pi^2 C}$. The variance is $\sim 0.01$ for $40$ nm DNA and $\sim 0.05$ for $160$ nm DNA. The writhe distribution based on energy Eq.~(\ref{energy}) was obtained by the normalized histogram of the recorded writhe numbers using a bin size of $0.01$. The normalized frequency is then divided by the bin size to obtain the density distribution function $\rho_{wr}(\textrm{Wr})$. Fig. \ref{fig_writhe}(a-d) shows $\rho_{wr}(\textrm{Wr})$ for DNAs of four different sizes that are subjected to various excitation energies. Clearly, the excitations in general lead to larger fluctuations of $\textrm{Wr}$.

\begin{figure}
\includegraphics[width=6cm]{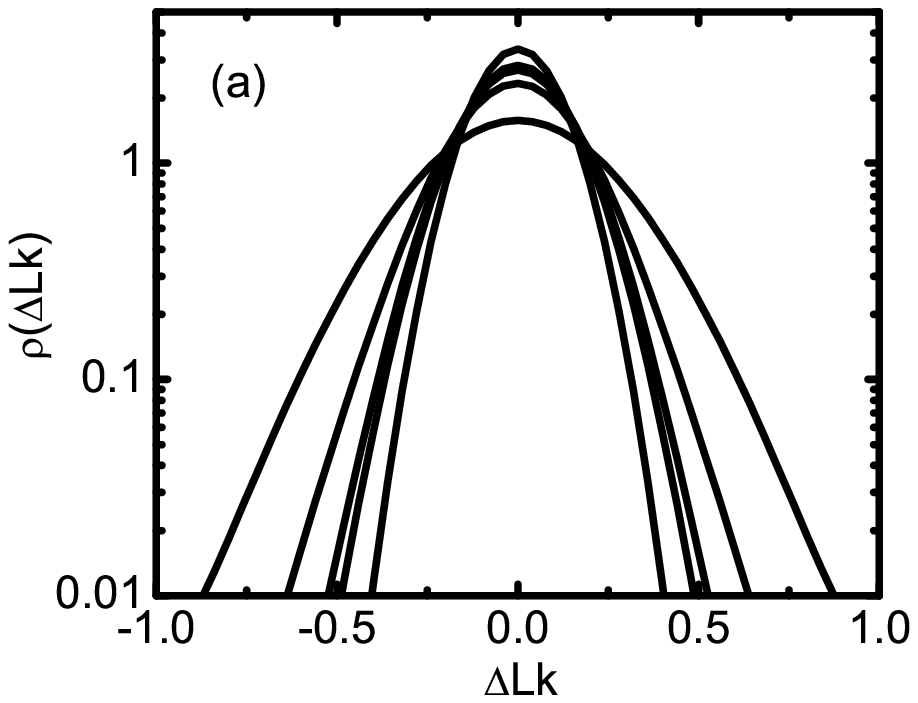}% Here is how to import EPS art
\includegraphics[width=6cm]{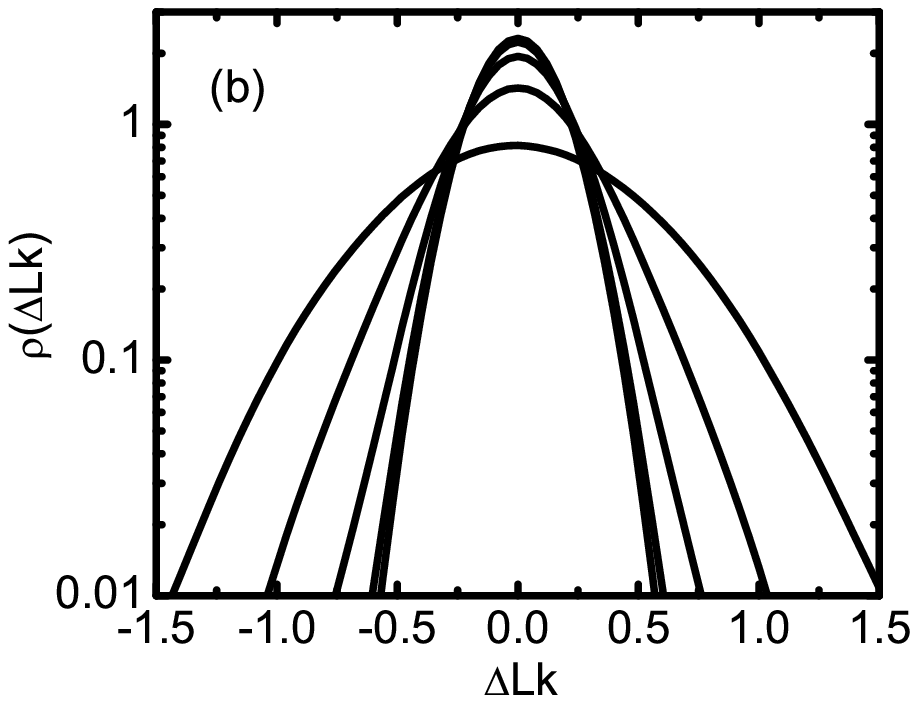}
\includegraphics[width=6cm]{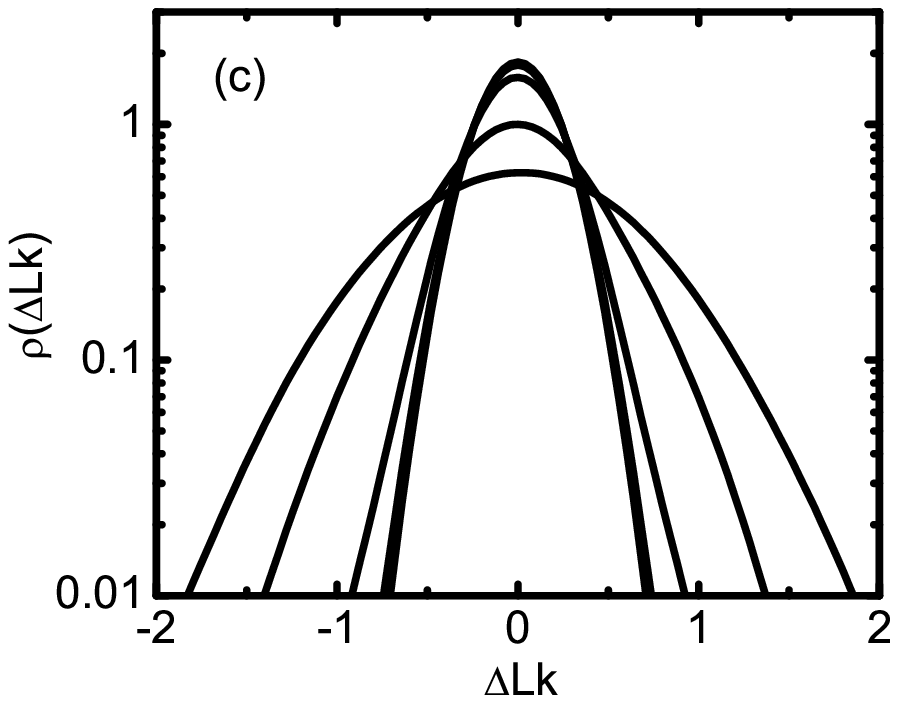}
\includegraphics[width=6cm]{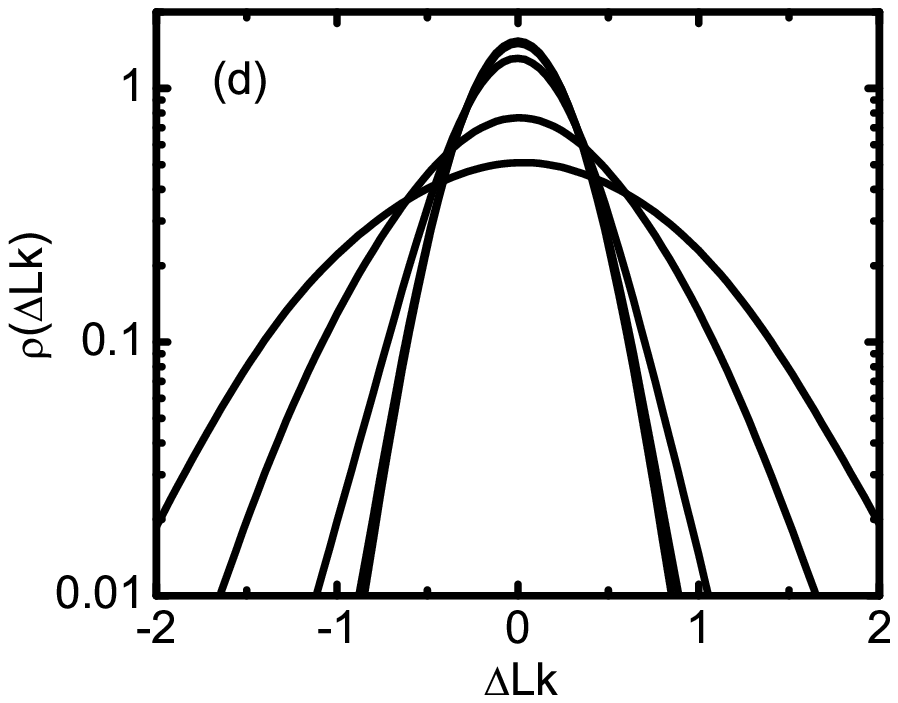}
\caption{Linking number distributions of DNA minicircles based on Eq. (\ref{energy}) of DNA minicircles of sizes $40$ nm (a), $80$ nm (b), $120$ nm (c), and $160$ nm (d), respectively. For each figure, from outside to inside, the excitation energies are $4$, $6$, $8$, $10$ $k_B T$, and $\infty$, respectively.
}
\label{fig_lk}
\end{figure}

Once $\rho_{wr}(\textrm{Wr})$ is obtained by simulation, $\rho(\Delta \textrm{Lk})$ is obtained by numerical convolutions of $\rho_{wr}(\textrm{Wr})$ and $\rho_{tw}(\Delta \textrm{Tw})$. The results are summarized in Fig. \ref{fig_lk} (a-d). Obviously, the widths of the distributions are noticeably increased by the excitations of defects. The variances of the distributions of $\rho(\Delta \textrm{Lk})$ are summarized in Table \ref{tab_lk1} (the left number in each bracket). From the table, the variances can change by a few folds in the range of the excitation energy studied. Table I shows that for DNAs $\geq 80$ nm and at the excitation energy $\mu\geq 10$ $k_B T$, the variances are very close to the predictions by the traditional WLC model. This is in agreement with the previous DNA cyclization experiments done at temperatures 20 - 30 degrees Celsius \cite{Shore1983,Cloutier2004}, and with the DNA linking number distribution measurements for DNAs larger than $200$ bp done at temperatures 4 - 37 degrees Celsius \cite{Horowitz1984,Depew1975,Rybenkov1997}. For DNA of 40 nm, the variance at $\mu= 10$ $k_B T$ is twice of that predicted by WLC model. However, the variance is perhaps too small to be measured in experiment. As shown in \cite{Horowitz1984}, a linking number variance as small as 0.03 can be experimentally measured. As such, the effects of the defect excitation can be investigated for 80 - 160 nm DNAs at $\mu\leq 8$ $k_B T$. As shown in Table I, under this condition the variances of the linking numbers are apparently larger than that predicted by the WLC model.

%yan: new section
\section{Effects of reduced twist rigidity and unwinding of the defect}
In the previous section, we show that with excitation of flexible defects, the flexibility of the defects increases the variance of the writhe distribution, and hence the variance of the linking number distribution. The mean of the linking number distribution is still zero, therefore, the major information of the effects of the defect is only reflected by the changes in the variance of linking number distribution. The variance of $\Delta \textrm{Lk}$ may be further increased if the defect is accompanied by reduced twist rigidity. It is straightforward to show that $\rho_{tw}(\Delta \textrm{Tw})$ is still a Gaussian with a variance :
\begin{equation}\label{variancetw2}
\sigma_{tw}^2 = \frac{1}{4 \pi^2} \left(\frac{(N-n)b}{C} + \frac{nb}{C'}\right),
\end{equation}
where $n$ is the number of defects in the DNA, and $C'$ is the twist persistence length of the defect. In this section, we consider a reduction of the twist persistence length of $C' = C/4$.

In addition to changes in the bending rigidity and twist rigidity, the defect may also lead to unwinding of DNA. Obviously, the underwinding may affect both the mean and variance of the linking number distribution. In B-form DNA, the relaxed twist angle per base pair step is about ${2\pi \over 10.5}$. A defected segment of $1$ nm ($3$ bp) is likely to destroy the local geometry constraint of the twist, leading to unwinding by an angle $\phi$, hence a relaxed twist angle per segment is about ${6\pi \over 10.5} - \phi$. The previous sections considered a limiting case where $\phi=0$. In this section, we consider the effects of $\phi > 0$ up to ${6\pi \over 10.5}$. 

Under these conditions, $\rho_{tw}$ and $\rho_{wr}$ are obviously no longer independent from each other since they both depend on the excitation of defect. Therefore, the linking number distribution is given by equation
\begin{equation}\label{eq.lk_n}
 \begin{array}{ccl}
 \rho(\Delta \textrm{Lk}) &=& \sum\limits_n P_n \rho_{lk}(\Delta \textrm{Lk}|n)\\
   &=& \sum\limits_n P_n \int d \textrm{Wr} \rho_{tw}(\Delta \textrm{Lk} - \textrm{Wr}|n) \rho_{wr}(\textrm{Wr}|n),\\
 \end{array}
\end{equation}
where $P_n$ is the probability of the DNA to contain $n$ defects, and $\rho(x|n)$, where $x$ refers to $\Delta \textrm{Lk}$ or $\textrm{Wr}$ or $\Delta \textrm{Tw} = \Delta \textrm{Lk} - \textrm{Wr}$, is the corresponding conditional distribution. 

Numerically the linking number distribution can be conveniently calculated directly from MC simulation by equation
\begin{equation}\label{eq.lk_mc}
\rho(\Delta \textrm{Lk}) = \frac{1}{M}\sum\limits_{i=1}^M \rho_{tw,n_i}(\Delta \textrm{Lk} - \textrm{Wr}_i),\\
\end{equation}
where $M$ is the total number of conformations sampled in the MC simulation. At each step of the simulation, the writhe number of the conformation is computed and substituted in the argument of $\rho_{tw}$. The expression of $\rho_{tw,n_i}$ is a Gaussian function with a variance determined by Eq.~(\ref{variancetw2}) and average value $-n_i \phi$, where $n_i$ is the number of defects excited in the $i^{th}$ conformation.

We first look at the combined effects of ($A'= 1$ nm, $C' =C/4 = 18.75$ nm, and $\phi = 0$) (Table \ref{tab_lk1}, right number in each bracket). It shows that the reduction in twist rigidity of $C'=C/4$ does not significantly increase the linking number variance: for $\mu \geq 10$ $k_B T$, the variances are nearly the same as those of $C'=C$; For $\mu < 10$ $k_B T$, the variances are only slightly larger than those of $C'=C$. This is not surprising: it simply means that, in the presence of a flexible defect, the major contribution to the changes in the linking number distribution is from the changes in writhe distribution. 

\begin{figure}
\includegraphics[width=6cm]{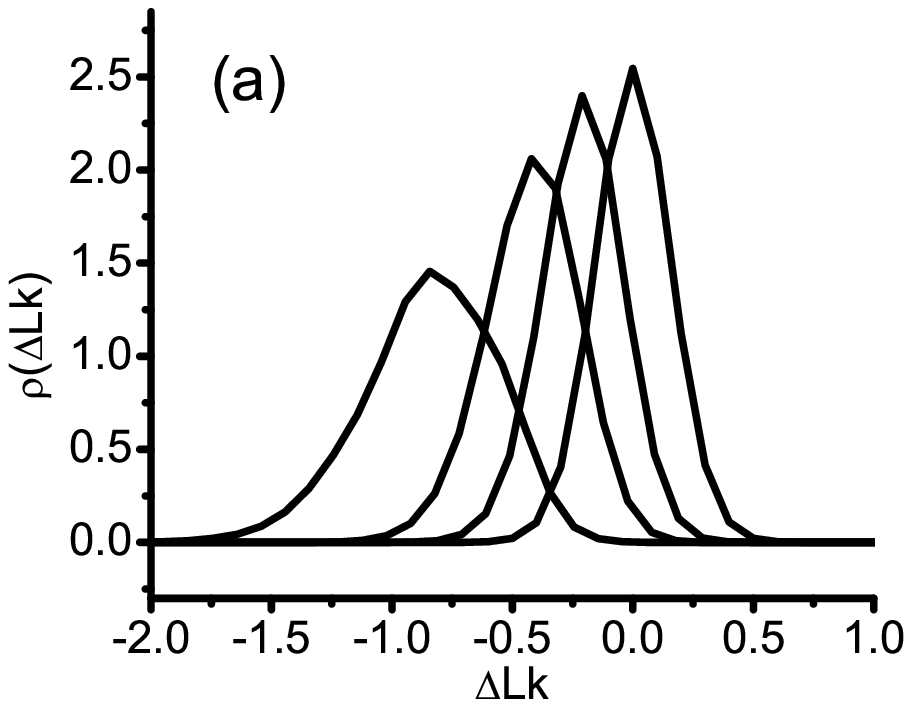}% Here is how to import EPS art
\includegraphics[width=6cm]{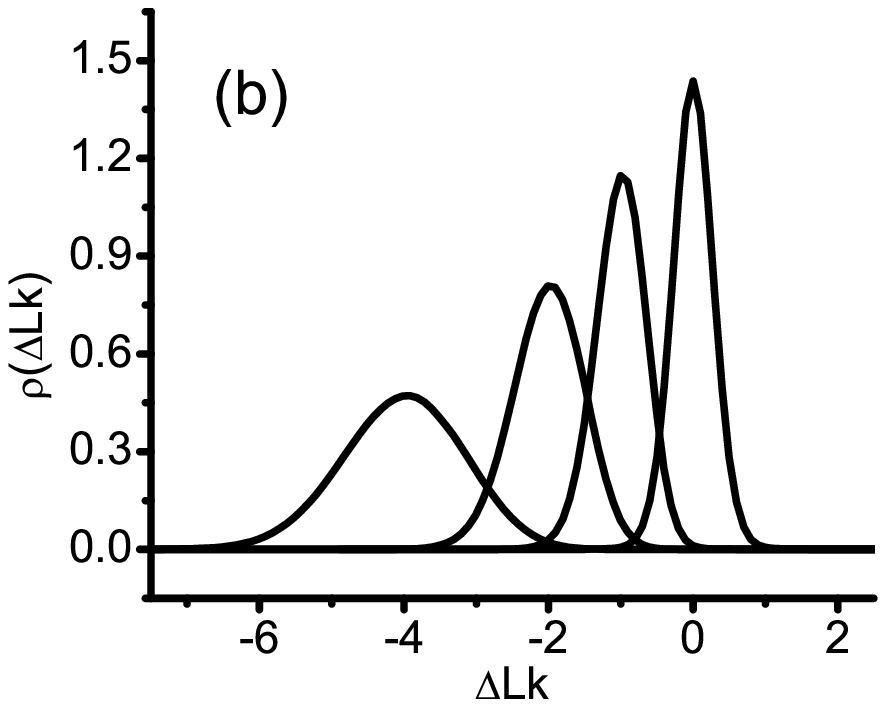}
\includegraphics[width=6cm]{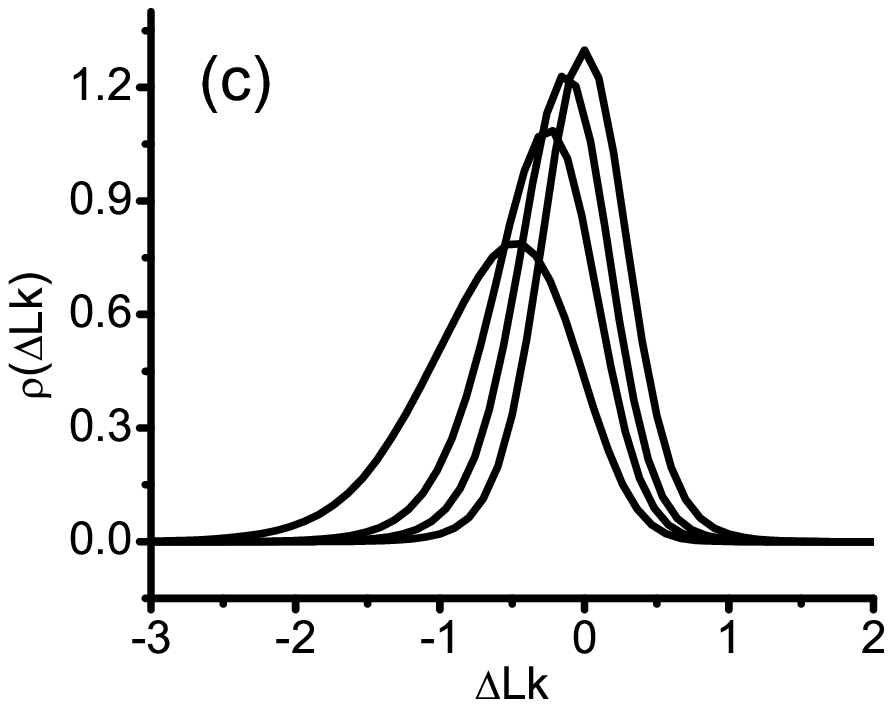}% Here is how to import EPS art
\includegraphics[width=6cm]{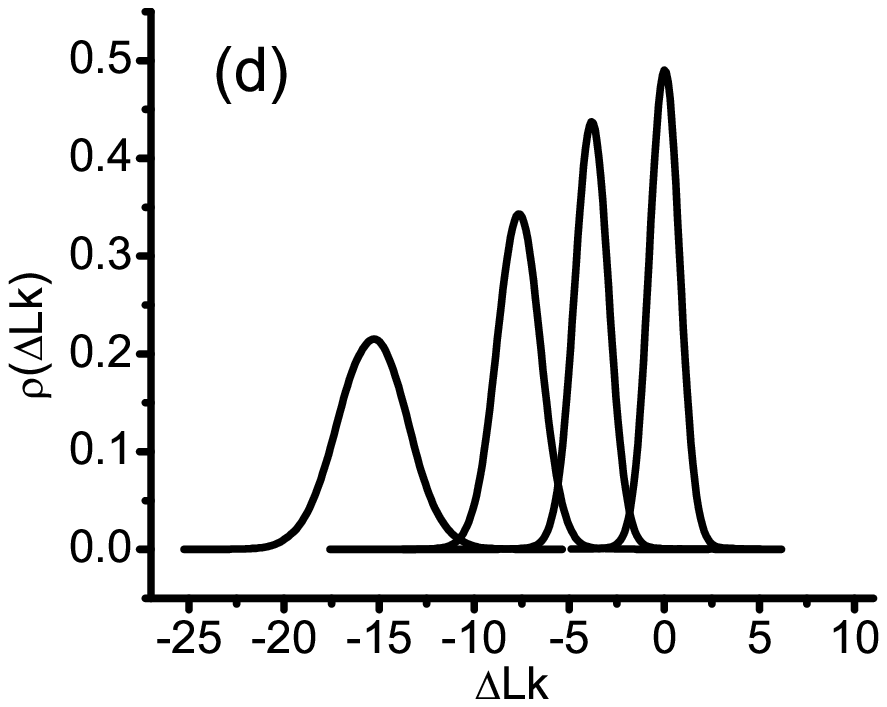}
\caption{Linking number distributions of DNA minicircles. (a) $40$ nm, $\mu=8$ $k_B T$; (b) $40$ nm, $\mu=4$ $k_B T$; (c) $160$ nm, $\mu=8$ $k_B T$; (d) $160$ nm, $\mu=4$ $k_B T$. In each figure, from right to left, the unwinding angle per defect is $\phi=0$, $26$, $51$, and $103$ degrees. $C' = C/4 = 18.75$ nm was used in the calculation.
}
\label{fig_lk_unwind}
\end{figure}

In contrast to the mild effect of $C'$, we expect a dramatic effect of $\phi > 0$. Firstly, it may change the mean of the linking number since unwinding leads to an offset of the relaxed twist angle per segment; secondly, it may further increase the linking number distribution since the averages of linking number are dependent on number of defects in the DNA, and the number of defects is not a constant. 
Fig. \ref{fig_lk_unwind} shows the linking number distribution of $40$ and $160$ nm circular DNAs with excitation energy $\mu=4$ and $8$ $k_B T$. As expected, with increasing of unwinding angle, linking number drifts away from the original equilibrium value of zero, and the distribution becomes wider. At the lower excitation energy $\mu=4$ $k_B T$, the average linking number drifts to much lower values due to the larger number of excited defects.

\section{Discussions}
To relate the excitation energy $\mu$ to temperature, the candidates of the possible defects should be discussed. Relevant to the DNA base pair stability, we consider two candidates: 1) the unstacking between two adjacent DNA base pairs, and 2) the melting of one or a few DNA base pairs. It was found recently that the temperature sensitive stacking interaction is the main stabilizing factor in DNA double helix, while the $A\cdot T$ pairing is always unstable and the $G\cdot C$ pairing does not contribute to the stability. In addition, the base pairing was found independent of temperature \cite{Yakovchuk2006}. However, this does not mean that melting does not cost more energy than unstacking. Instead, there is an additional large energy penalty for melting one base pair inside an DNA double helix. This energy penalty is related to a temperature and sequence insensitive ``ring factor'' $\xi\approx 10^{-3}$, corresponding to an energy penalty $\approx 6.6$ $k_B T$ \cite{Krueger2006}. The real melting energy is roughly the unstacking energy plus this large energy penalty. 

On the other hand, the excitation energy $\mu$ in our model is defined for a given DNA conformation, while in a bulk measurement all the possible conformations contribute. Therefore, to relate it to the unstacking or melting energy $\bar{\mu}$ measured in the bulk experiments, one needs to consider the conformational entropy gain resulted from the defect excitation. According to the excitation model Eq.~(\ref{WLC_bubble}), for an unconstraint DNA, the probability of a segment to be defected is $p = (1+ {i_0(a)\over i_0(a')}
e^{\mu} e^{(a'-a)})^{-1}$, where $i_0$ is the zero$^{th}$ order modified spherical Bessel function of the first kind \cite{Yan2005}. On the other hand, for a two states lattice model it should be $p = \left(1+e^{\bar{\mu}}\right)^{-1}$. Direct comparison gives $\bar{\mu} = \mu - {a\over a'}\ln {1-e^{-2a'}\over 1-e^{-2a}}$ (please not this relation can be derived by comparison between the chain Hamiltonian of the defect excitation model and that of the two state lattice model \cite{Sivak2008}). Using our parameters $a=50$, and $a'=1$, $\bar{\mu} \approx \mu - 3.7$ $k_B T$. 
Thus, the energies $\mu = (4, 6, 8, 10, \infty)$ correspond to the bulk values $\bar{\mu}=(0.3, 2.3, 4.3, 6.3, \infty)$.

As shown in \cite{Yakovchuk2006}, the energy contributions from stacking to DNA base pair stability are 1.6 - 3.3 $k_B T$ for $G\cdot C$ base pair and 1.1 - 2.7 $k_B T$ for $A\cdot T$ base pair, respectively, in the temperature range 20 - 55 degrees Celsius. Hence, the melting energy of one base pair in the temperature range is 8.2 - 9.9 $k_B T$ for $G\cdot C$ base pair and 4.4 - 9.3 $k_B T$ for $A\cdot T$ base pair, respectively. Obviously, the melting energies measured in bulk experiments are larger than our excitation energies. Therefore, one should not expect to see much melting related changes in linking number distribution. However, according to the data, our excitation energies are sufficient to unstack neighboring base pairs. Although it is unclear whether the unstacking defect is a flexible defect, it is reasonable to think that it may lead to the formation of a ``kink''. Since a flexible defect in a small circular DNA is presented as a ``kink'', the prediction based on flexible defect excitation should also describe the effects of kink excitation on the linking number distributions. In addition, unstacking a base pair step is likely associated with local unwinding, so the discussion of the effect of unwinding may also be relevant.

As shown in our computation, the excitation of defects can change the linking number distribution through three ways: 1) increase the variance of the writhe distribution by reduced local bending rigidity (equivalently, formation of local kink), therefore increase the linking number variance; 2) increase the variance of the twist distribution by reduced local twist rigidity, therefore increase the linking number variance, and 3) reduce the average and increase the variance of the twist by local unwinding, therefore drift and widen the distribution of linking number. At low excitation energies (i.e., high temperatures), 1) and 3) dominate. In addition to study DNA base pair stability, these predictions may be useful in other applications. For example, one can use it to study non-specific interactions between DNA and DNA-distorting proteins, where the protein binding can be treated as the excitation of ``kink'' or ``unwinding'' defect. The change in the linking number distribution can then be related to the binding affinity, the bending angle, and unwinding angle induced by protein binding. 

In summary, we have predicted the temperature dependence of linking number distributions of small circular DNAs in the temperature range 20 - 55 degrees Celsius. This temperature range is convenient for experimental studies. For temperatures below 37 degrees Celsius, one can use the T4 ligase to close DNAs into loops. For temperatures higher than 37 degrees Celsius, one can use Taq DNA ligase (it is known active within the range 45 - 65 degrees Celsius) to close the DNAs into loops. Once the linear DNA is looped, its linking number is fixed, so the linking number distribution can be measured at any convenient temperatures. Although our predictions are based on excitations of flexible defects, they should also describe the effects of excitations of ``kinks''. The most likely candidate of the defect is the unstacking of one base pair step which is presumably a ``kink''. For another candidate, the melting of one or a few DNA base pair, costs a much higher energy, therefore it is less likely to be excited. We expect to see increase in the variance of linking number distribution at high temperatures, which is likely associated with drift in the mean. We also emphasize an potential application for studies of DNA-protein interactions by looking at the change in linking number distribution due to protein binding. 

Finally, we note that our predictions of temperature dependence of linking number distribution of short DNAs should not be understood as accurate quantitative predictions. This is because the energy model Eq. (\ref{energy}) used in the simulation contains two parameters, so there are certain degrees of freedom to choose the parameters. The choice of the parameters $(A', \mu)$ should agree with the predictions by WLC model for DNA not sharply bent and at room temperature. From Table \ref{tab_lk1}, we see that for $\mu > 10$ $k_B T$, the linking number variances converge to those predicted by the WLC model. In addition, previously we showed that the choices of $A'=1$ nm and $\mu=11$ $k_B T$ agree with the force-extension curve measured in single-DNA stretching experiments, and the ``j-factor'' measurement of $> 230$ bp DNAs \cite{Yan2004,Yan2005}. Furthermore, this set of parameters also explain the ``j-factor'' measurement of $\sim 100$ bp DNA at 30 degrees Celsius reported in \cite{Cloutier2004,Cloutier2005} (but note that another ``j-factor'' measurement of $\sim 100$ bp DNA done at a lower temperature 21 degrees Celsius agreed with the prediction by the traditional WLC model \cite{Du2005}). If one chooses a larger $A'$, then $\mu$ becomes correspondingly smaller in order to be in agreement with the experiments. However, our predictions of the temperature dependence of linking number distribution of short DNAs should still qualitatively hold.

\section*{Acknowledgement}
This research was supported by the Ministry of Education of Singapore through Grants No. R144000143112 and No. R144000171101.
JY thanks John Marko for valuable discussions.
YL thanks the Foundation for the Visiting PhD Candidate of the Chinese Academy of Science.

\newpage %Just because of unusual number of tables stacked at end
%\bibliography{yanhui}% Produces the bibliography via BibTeX.

\newpage

\begin{table}
\caption{Variances of the linking number distributions of DNA minicircles of sizes 40 nm, 80 nm, 120 nm, and 160 nm, which are subject to excitation energies (in $k_B T$ unit) $4$, $6$, $8$, $10$, and $\infty$. In each bracket, the left number is computed using $C'=C=75$ nm, while the right number was computed using $C' = C/4 = 18.75$ nm.
}\label{tab_lk1}
\begin{ruledtabular}
\begin{tabular}{cccccc}
Size (nm) &$\mu=4$ $k_B T$ &$\mu=6$ $k_B T$ &$\mu=8$ $k_B T$ &$\mu=10$ $k_B T$ &$\infty$\\
\hline
40 &(0.07,0.08) &(0.03,0.04) &(0.02,0.03) &(0.02,0.02) &(0.01,0.01) \\
80 &(0.25,0.26) &(0.09,0.10) &(0.05,0.05) &(0.03,0.03) &(0.03,0.03) \\
120 &(0.39,0.47) &(0.19,0.19) &(0.07,0.07) &(0.05,0.05) &(0.05,0.05) \\
160 &(0.55,0.68) &(0.30,0.31) &(0.10,0.11) &(0.07,0.07) &(0.07,0.07) \\
\end{tabular}
\end{ruledtabular}
\end{table}

\end{document}